# High flux coherent supercontinuum soft X-ray source driven by a single-stage 10 mJ, kHz, Ti:sapphire laser amplifier

Chengyuan Ding[1,*], Wei Xiong[1], Tingting Fan[1], Daniel D. Hickstein[1], Tenio Popmintchev[1], Xiaoshi Zhang[2], Mike Walls[2], Margaret M. Murnane[1], and Henry C. Kapteyn[1]

[1]*Department of Physics, University of Colorado and JILA, National Institute of Standards and Technology and University of Colorado, Boulder, CO 80309-0440, USA*
[2]*Kapteyn-Murnane Laboratories, 1855 S. 57th Court, Boulder, CO 80301, USA*
*chengyuan.ding@colorado.edu*

**Abstract:** We demonstrate the highest flux tabletop source of coherent soft X-rays to date, driven by a single-stage 10 mJ Ti:sapphire regenerative amplifier at 1 kHz. We first down-convert the laser to 1.3 µm using a parametric amplifier, before up-converting it to soft X-rays using high harmonic generation in a high-pressure, phase matched, hollow waveguide geometry. The resulting optimally phase matched broadband spectrum extends to 200 eV, with a soft X-ray photon flux of $> 10^6$ photons/pulse/1% bandwidth at 1 kHz, corresponding to $> 10^9$ photons/s/1% bandwidth, or approximately a three order-of-magnitude increase compared with past work. Finally, using this broad bandwidth X-ray source, we demonstrate X-ray absorption spectroscopy of multiple elements and transitions in molecules in a single spectrum, with a spectral resolution of 0.25 eV, and with the ability to resolve the near edge fine structure.

**Introduction**

The generation of coherent beams in the x-ray region of the spectrum has been pursued for many decades, because of the potential to combine the spatial and temporal coherence of lasers with the atomic-level spectroscopic and spatial resolution characteristics of x-rays. Partially coherent X-ray beams can be generated from large-scale synchrotron facilities, and more recently using (also large-scale) X-ray free electron lasers (XFELs) that generate high per-pulse energies and spatial coherence. High-order harmonic generation (HHG), on the other hand, coherent upconverts light from a tabletop femtosecond laser, while retaining full spatial and temporal coherence when implemented in a phase matched geometry [1-5]. Thus HHG provides an advanced light source with capabilities complementary to facility-scale sources. By producing x-ray pulses with durations of <100 attoseconds [6-8] and wavelengths < 1 nm [3], HHG is opening up new research opportunities in molecular spectroscopy [9-12], materials science [13-19], as well as nanoscale coherent diffractive imaging [20, 21].

Most implementations of HHG to-date have used Ti:sapphire driving lasers at a wavelength around ~0.8 µm. In this case, phase matching considerations limit bright HHG to photon energies $\lesssim 100$ eV, which lie in the extreme ultraviolet (EUV) region of the spectrum. This limits applications of HHG to very thin samples and a relatively small number of elements which have absorption edges in the EUV. Other work extended HHG into the soft X-ray region using short 12 fs lasers pulses in a non-phase matched implementation, with a flux of $10^5$ photons/s in 1% bandwidth [22]. Fortunately, a recent breakthrough extended phase matching of HHG to much higher keV photon energies (i.e. *shorter* HHG wavelengths), by using *longer* wavelength mid-infrared driving lasers at wavelengths from 1.3 to 3.9 µm [3, 23-26]. However, these preliminary studies were done using lasers operating at 10 Hz repetition rates, with a total photon flux of $10^6$ photons per second in 1% bandwidth [25], which is insufficient for broad applications in spectroscopy or imaging. To fully harness the unique properties of soft X-ray HHG, kHz or higher repetition-rates are required in a phase matched geometry. This goal requires the development of stable, compact, high pulse-energy, ultrafast laser amplifiers that can generate multi-millijoule, < 10 cycle, mid-infrared laser pulses required for efficient HHG.

We note that since HHG is a highly-nonlinear process, it is very sensitive to fluctuations in peak laser intensity or beam pointing. Moreover, the efficiency of the HHG process scales rapidly with pulse energy, and requires >1 mJ pulse energy to obtain absorption-limited conversion efficiency in the soft X-ray region [3, 23-26]. To obtain this pulse energy in a femtosecond mid-infrared pulse, optical parametric amplification (OPA) [27] and optical parametric chirped-pulse amplification (OPCPA) [26] can be employed [3, 23-26, 28-30]. However, each approach has shortcomings for HHG. OPCPA systems are still experimental in this parameter range, and do not yet possess the stability, multi-mJ, kHz repetition rates, and reliability required for HHG application. Similarly, although Ti:sapphire laser pumped OPAs are a well-established technology, this approach requires a ~10 mJ pulse energy from the Ti:sapphire laser. The complexity of the traditional two-amplification-stage lasers required to pump an OPA greatly increases the cost and difficulty of implementing a table-top HHG soft X-ray source reliably on a daily basis.

In this letter, we demonstrate the highest pulse energy (10 mJ), single-stage, ultrafast (45 fs) Ti:sapphire amplifier to date, with a repetition rate of 1 kHz. We then use this laser to pump an optical parametric amplifier system and generate 1.3 µm, 30 fs pulses with sufficient energy (2 mJ) for optimally-efficient, phase matched HHG conversion. This allows us to demonstrate the highest flux, soft X-ray HHG source to date with > $10^6$ photons/pulse/1% bandwidth at 1 kHz (corresponding to > $10^9$ photons/s/1% bandwidth) in a broadband, continuum, spectrum extending to 200 eV. This photon flux represents an approximately 3 orders-of-magnitude increase compared with past work, which used lower repetition rates, or non-phase matched implementations [24, 29]. Finally, using this unique bright *supercontinuum* HHG soft X-ray source, we demonstrate soft X-ray absorption spectroscopy of multiple elements and transitions in molecules simultaneously, with 0.25 eV spectral resolution, and with the ability to resolve near edge fine structure with high fidelity. This work represents a viable approach for implementing time-resolved HHG soft X-ray absorption spectroscopy studies in a new region of the spectrum, with femtosecond to attosecond time resolution. Moreover, the ability to capture near edge X-ray absorption fine structure (NEXAFS) data simultaneously has significant implications for molecular and material spectroscopies [31].

**Single stage, 10 mJ, femtosecond laser system**

We first discuss the 10 mJ, single-stage, 1 kHz regenerative Ti:sapphire amplifier system (KMLabs Wyvern HE™) and high-efficiency OPA that generates high quality 1.3 µm driving pulses for soft X-ray HHG. The maximum output pulse energy obtainable from an ultrafast

Ti:sapphire regenerative amplifier is primarily limited by two factors: thermal lensing [32] and optical damage. By employing closed-cycle cryogenic cooling of the Ti:sapphire crystal, the temperature of the Ti:sapphire crystal can be maintained below 70 K, even when pumped with 50 W of 532 nm light. This dramatically (i.e. by >>100x) reduces the thermal lens present in the laser crystal [33]. Damage of the optical components by the intense laser pulse is overcome through a new design of the regenerative cavity. By increasing the cavity length and temporally stretching the seed pulse to 150 ps using with high groove density gratings (1800 lines/mm), the amplified pulse fluence of 1.0 J cm$^{-2}$ at 150 ps is kept below the damage threshold of commercial broadband optics (typically ≈ 1.5 J cm$^{-2}$ at 150 ps). The maximum output (compressed) pulse energy is 10.6 mJ when pumped with 50 mJ of 527 nm light from a Nd:YLF pump laser. Since the efficiency of the compression gratings is 70%, this represents an optical-to-optical efficiency of the regenerative amplifier of ~30% before compression. To our knowledge, this represents at least a 25% improvement in the pulse energy obtainable from a single-stage 1 kHz Ti-Sapphire amplifier [34]. This increased pulse energy is critical since the OPA and HHG processes are non-linear: therefore the conversion efficiency into the mid-infrared and then to soft X-rays increases dramatically with increasing energy of the Ti:sapphire driving pulse.

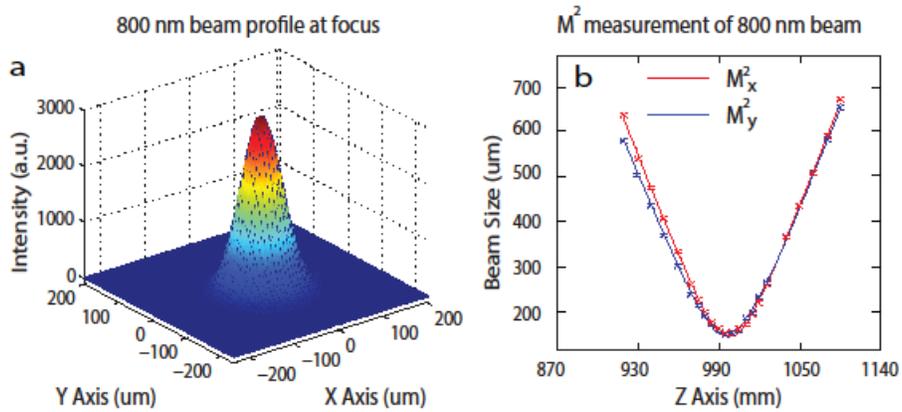

Fig. 1. (a) Focal spot of the single-stage laser amplifier output (KMLabs Wyvern HE$^{TM}$), using a 1 m focal length lens; (b) M$^2$ characterization of the 0.8 µm amplifier output. Data taken at 10 mJ, 1 kHz.

The OPA and HHG processes are also extremely sensitive to the beam mode and the pulse duration. The 10 mJ single-stage regenerative Ti:sapphire amplifier has a nearly perfect Gaussian mode output (see Fig. 1(a)), with M$^2$ [35] values for both x-axis and y-axis of ~1.1, which means almost all the pulse energy is in the lowest order Gaussian mode. This feature ensures good conversion efficiency and output mode for the OPA [36], which in turn ensures high HHG efficiency (since a good OPA mode efficiently couples into a waveguide and facilitates good phase matching). The pulse duration after the compressor is ≲45 fs, and is primarily limited by the bandwidth of the thin film polarizer (TFP) inside the amplifier cavity. Unlike a conventional regenerative cavity, the dispersion introduced by the material in the regenerative cavity is not the limiting factor for the pulse duration because only 12 passes through the Ti:sapphire crystal are needed to reach the full output power (15 mJ before compression), which is approximately the same number of passes used in a typical multipass ti:sapphire amplifier, but with much higher efficiency. In contrast, a typical regenerative amplifier requires 20-40 passes. In the future, a shorter pulse can be achieved when broader bandwidth, higher-damage-threshold, TFPs becomes available.

To obtain high-quality 1.3 µm pulses to drive soft X-ray HHG, we use a homebuilt 3-stage OPA that achieves greater than 40% conversion of the 0.8 µm light into 1.3 µm (signal) and 2 µm (idler) beams. For these measurements, the total pulse energy used to drive the OPA was 8.5 mJ. In the first stage, 1% of the pulse energy is used for white light generation and the first stage pump. The second stage uses 9% of the pulse energy as the pre-amplification pump, while 90% of the pulse energy is used to pump the final stage. The output OPA signal pulse energy is 2.25 mJ, with a spectral full-width of greater than 200 nm centered at 1.3 μm. The pulse energy of the idler is 1.4 mJ, giving a total conversion efficiency in the OPA from pump to signal plus idler of approximately 43%. Additionally, the broad spectrum of the 1.3 μm light originates from white light generation, and supports an ultrashort pulse duration of 30 fs (characterized by a second harmonic FROG measurement). As expected in parametric down-conversion of multi-cycle ultrafast pulses, this signal pulse duration is considerably shorter than the 45 fs 0.8 µm pump pulse.

**High flux harmonics in the soft X-ray region driven by 1.3 µm OPA beams**

Bright soft X-ray harmonics are generated by focusing the 2 mJ pulse energy, 1.3 µm, OPA signal output into a gas-filled hollow waveguide (length 1 cm and diameter 150 μm) using an f = 25 cm lens. The coupling efficiency is more than 50% even with high pressure gas in the waveguide, so that the driving pulse peak intensity inside of the waveguide can reach $5 \times 10^{14}$ W/cm$^2$, and the coupled energy is all in $EH_{11}$ mode. The waveguide geometry is essential for achieving fully phase-matched high flux soft X-ray harmonics for two reasons. First, use of a guided mode ensures nearly plane-wave propagation of the mid-infrared driving beam, to maintain a flat uniform phase and high peak intensity over a large interaction distance. And second, the waveguide confines the gas, allowing high gas pressures (>1 atm) to be used, which is critical for achieving absorption-limited phase matched HHG flux for longer driving wavelengths. This high-pressure gas medium, together with an abrupt transition to vacuum, becomes increasingly important as the HHG driving laser wavelength is increased, because the phase matching pressure (multiple atmospheres) is higher than for 0.8 µm driven HHG. Moreover, since the single-atom HHG yield drops, more emitters are needed to obtain high flux harmonics.

To generate bright soft X-rays, the waveguide is filled with Argon (Ar) or Neon (Ne). Since each noble gas has a different refractive index and ionization potential, each gas optimizes at a different phase matching pressures and laser intensity, and also extends efficient HHG to some maximum phase matched photon energy cutoff. By controlling the gas pressure inside the waveguide and the 1.3 µm pulse energy before the waveguide, we can optimize the HHG flux at the phase matching pressure, as shown in Fig. 2. For a 1 cm long, 150 µm diameter waveguide and 1.3 µm driving lasers, the optimized pulse energies, phase matching pressures and phase matching cutoff photon energies are 0.95 mJ, 700 torr and 110 eV for Ar, and 1.6 mJ, 1300 torr, and 200 eV for Ne. By accounting for the absorption of the metal filters, the efficiencies of the spectrometer (Hettrick Scientific), and the quantum efficiency of the CCD camera (Andor, Inc.), the photon flux of the harmonics is estimated at more than $10^6$ photons/pulse in 1% bandwidth up to 200 eV (Fig. 2), corresponding to $10^9$ photons/s in 1% bandwidth at kHz repetition rates. This photon flux represents an approximately 3 orders of magnitude increase compared with past work (which used lower repetition rates or non-phase matching geometry) [24, 29]. Additionally, a long term stability test demonstrates that the HHG flux is stable over many hours, limited primarily by beam pointing drift (when no beam pointing stabilization is used).

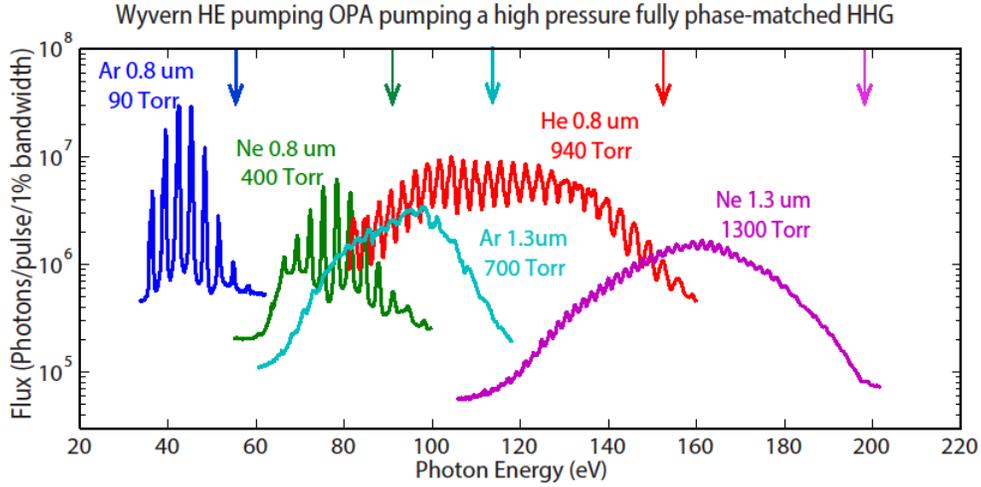

Fig. 2. Flux characterization and comparison of the optimized fully phase matched HHG emission from a 1 cm long, 150 μm diameter, waveguide driven by 1.3 μm and 0.8 μm light in various noble gases (Ar, Ne, and He). The vertical arrows indicate the maximum predicted phase matched HHG energy cutoff for each gas and laser wavelength. The HHG flux obtained using a 1.3 μm driving laser is comparable to that achieved using 0.8 µm.

Since numerous experiments have been implemented utilizing 0.8 µm-driven EUV harmonics, we compared the optimized fully phase matched HHG flux driven by 0.8 µm light with the HHG flux driven by 1.3 µm light. We used the same waveguide for comparison, since the geometry and quality of the waveguide might affect the harmonic yield. The pulse energy used for the 0.8 µm-driven HHG from Ar, Ne, and He is 0.40 mJ, 1.5 mJ and 2.2 mJ, respectively. We find that at 100 eV photon energy, the 1.3-µm-driven HHG flux is within a factor of two of the flux of 0.8 µm-driven HHG, while above 150 eV, the 1.3 µm HHG flux becomes much more efficient. This shows that many applications based on EUV harmonics can now be extended to the soft X-ray region. Although our laser/OPA system did not provide sufficient pulse energy to generate fully optimized phase-matched HHG flux from He, we were in-fact able to generate coherent light up to ~300 eV from He (See Fig. 3(a)). However, the HHG flux obtained was significantly lower (by ~100x) than the data of Fig. 2, which still represents a record HHG flux of $10^7$ photons/s in 1% bandwidth at 300 eV, at kHz repetition rates. We predict that a further modest increase in pulse energy (by ≲ 2x) will allow us to generate flux comparable to that of Fig. 2 up to the water window using He as the nonlinear medium [24].

The high-flux, broadband HHG spectra provide a unique opportunity to measure soft X-ray absorption spectra from multiple orbitals of an element or multiple elements simultaneously. Additionally, the kHz repetition rate ensures high quality absorption spectra in a short data acquisition time, which is important for time-resolved experiments. By passing the harmonics through a 2 mm long, 200 μm diameter, cylindrical sample cell, the absorption signal from the sample can be directly observed from the transmitted HHG spectrum (Fig. 3(a)).

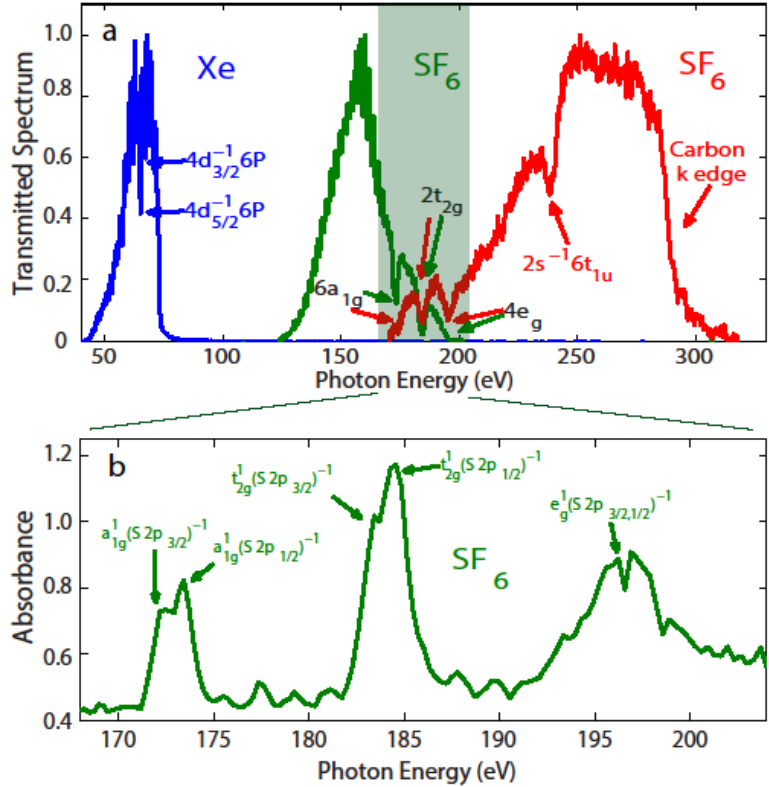

Fig. 3. (a) Normalized raw data for 1.3 μm-driven HHG from Ar (blue), Ne (green), and He (red) after transmission through a 2 mm long sample cell of Xe (HHG from Ar), and $SF_6$ (HHG from Ne and He). Note that the apparent gap between 70 eV and 100 eV is because of Al filters used to block the fundamental light—full tuning across this range can be achieved by switching to other filters as shown in Fig. 2. (b) NEXAFS spectrum of $SF_6$ Sulfur L-edge 2p orbital.

For $SF_6$ gas, using harmonics generated from Ne, we simultaneously observe three Sulfur 2p inner well resonance absorption peaks [37] $6a_{1g}$, $2t_{2g}$, $4e_g$, as well as $2t_{2g}$ and $4e_g$ shape resonances. When the $SF_6$ gas transimssion is measured using He HHG, one additional Sulfur 2s Fano-type resonance absorption peak [38] $2s^{-1} 6t_{1u}$ appears in the spectrum. The absorption peak $6a_{1g}$, which is not obvious in the spectrum of the harmonics from Ne, is much sharper in the He HHG spectrum. Note that in the He data, an absorption at a photon energy of 285eV originates from the carbon (C) K-edge, due to C contamination on the spectrometer optics. Thus, we can measure the absorption spectrum of multiple elements simultaneously.

Figure 3(a) also shows the 4d giant resonance absorption in Xe [39] at high spectral resolution using harmonics from Ar gas, which clearly resolves absorption from different spin states. This demonstrates element specificity and sensitivity to electronic structure over a very broad ~250 eV spectral range in a tabletop setup. High quality absorption spectra can be obtained by taking the logrithiam of the ratio between the transmitted harmonic spectrum without and with the sample gas, for the same exposure time. Figure 3(b) shows the $SF_6$ Sulfur L-edge 2p orbital NEXAFS structure. Different transitions and spin states are clearly resolved and assigned using synchrotron data [37]. Although the energy resolution is currently 0.25 eV, limited by 160 g/mm groove density of the grating used in our setup, further increases in spectral resolution will be possible using a higher groove-density grating.

Finally, it is worth noting that with this spectral resolution, individual harmonic peaks (separated by 1.9eV) will be clearly resolvable. Thus, the harmonic spectrum using Ne gas is a true coherent broadband supercontinuum - an observation which is consistent with other recent work demonstrating that phase matched soft X-ray HHG driven by mid-infrared lasers naturally emerges as an isolated, attosecond-duration, burst [40].

**Conclusion**

In conclusion, we have demonstrated a unique 10 mJ kHz single-stage Ti-Sapphire regenerative amplifier design capable of generating the highest-flux coherent soft x-rays to date using a two-step frequency upconversion conversion process. Simplifying the driving laser into a single stage laser greatly enhanced the stability and mode quality of the pump, which facilitate its application for an OPA-driven HHG setup. The 1.3 μm light generated from OPA is coupled into a high-pressure gas-filled waveguide to generate the soft X-ray HHG beams. By properly controlling the gas pressure and pulse intensity to fully phase match the HHG process, we achieve a photon flux of more than $10^9$ photons/s/1% bandwidth up to 200 eV at kHz repetition rates, which is enhanced by $> 10^3$ compared with past work. High-quality soft X-ray absorption spectroscopy of multiple elements and transitions from multiple electronic and spin states is also demonstrated using this unique soft X-ray light source. The stability and brightness of this tabletop soft X-ray source indicate great potential for future time resolved molecular and material dynamics studies.

**Acknowledgements**

We gratefully acknowledge support from the DARPA PULSE program, a DOE AMOS grant, and facilities provided by the National Science Foundation Engineering Research Center in EUV Science and Technology. We also would like to gratefully acknowledge support from the DOE SBIR programs, grant numbers DE-SC0007469 and DE-SC0006514, at KMlabs, which supported the development of the high energy regenerative amplifier system and the hollow waveguide design used for soft X-ray generation, respectively.